\documentclass[twocolumn,showpacs,aps,prd,superscriptaddress,preprintnumbers,amsmath,amssymb,floatfix]{revtex4}
\usepackage{graphicx}
\usepackage{dcolumn}
\usepackage{bm}

\usepackage{epsfig}
\usepackage{afterpage}
\usepackage{psfrag}
\usepackage{url}

\newcommand{\ra}{\ensuremath{\rightarrow}}

\newcommand{\mumu}{\ensuremath{\mu^{+}\mu^{-}}}

\newcommand{\nbel}{\ensuremath{{\overline{\nu}}_{e}}}

\newcommand{\MeV}{\ensuremath{\mathrm{MeV}}}

\newcommand{\nut}   {\nu_\tau}

\newcommand{\qqbar} {$q \overline{q}$}

\newcommand{\en}{\ensuremath{e \nbel}}

\newcommand{\ifb}{\ensuremath{\mathrm{fb^{-1}}}}

\newcommand{\ARGUS}{\mbox{\scshape Argus}}

\newcommand{\bfl}{\begin{flushleft}}
\newcommand{\efl}{\end{flushleft}}
\newcommand{\bfr}{\begin{flushright}}
\newcommand{\efr}{\end{flushright}}
\newcommand{\bc}{\begin{center}}
\newcommand{\ec}{\end{center}}

\newcommand{\Like}{\ensuremath{\mathcal{L}}}
\newcommand{\Acp}{\ensuremath{\mathcal{A}_{CP}}}

\newcommand{\pizero}  {\ensuremath{\pi^{0}}}

\newcommand{\pipm}    {\ensuremath{\pi^{\pm}}}
\newcommand{\Kpm}     {\ensuremath{K^{\pm}}}
\newcommand{\pimp}    {\ensuremath{\pi^{\mp}}}

\newcommand{\K}      {\ensuremath{K}}
\newcommand{\Kstar} {\ensuremath{\K^{*}}}
\newcommand{\Kstarz} {\ensuremath{\K^{*0}}}

\newcommand{\Kstarpm} {\ensuremath{K^{*\pm}}}

\newcommand{\rhomp}   {\ensuremath{\rho^{\mp}}}
\newcommand{\rhopm}   {\ensuremath{\rho^{\pm}}}
\newcommand{\rhostar} {\ensuremath{\rho^{*}}}
\newcommand{\rhostarpm} {\ensuremath{\rho^{*\pm}}}

\newcommand{\Bpm}     {\ensuremath{B^{\pm}}}
\newcommand{\Bp}      {\ensuremath{B^{+}}}
\newcommand{\Bm}      {\ensuremath{B^{-}}}
\newcommand{\B}       {\ensuremath{B}}
\newcommand{\Bbar}    {\ensuremath{\bar{B}}}
\newcommand{\BpBm} {\ensuremath{B^{+}B^{-}}}

\newcommand{\BB}   {\ensuremath{B\bar{B}}}

\newcommand{\rhopi}   {\ensuremath{\rhopm \pizero}}

\newcommand{\Kstarpi} {\ensuremath{\Kstarpm \pizero}}

\newcommand{\Btopipipi}  {\ensuremath{\Bpm \ra \pipm \pizero \pizero}}
\newcommand{\Btorhopi}   {\ensuremath{\Bpm \ra \rhopi}}
\newcommand{\Btokstarpi} {\ensuremath{\Bpm \ra \Kstarpi}}
\newcommand{\Btorhostarpi} {\ensuremath{\Bpm \ra \rhostarpm\piz}}

\newcommand{\thetabsph} {\ensuremath{\theta_{\mathrm{Sph}}^{\mathrm{B}}}}  
\newcommand{\ANN}       {\ensuremath{\mathrm{A_{NN}}}}
\newcommand{\DeltaE}  {\ensuremath{\DeltaE}}
\newcommand{\mes}     {\ensuremath{m_{\mathrm{ES}}}}

\newcommand{\rhop} {\ensuremath{\rho^{+}}}
\newcommand{\rhom} {\ensuremath{\rho^{-}}}

\input{babarsym.tex}

\def\LUMI{211}  
\def\BBpairs{232}  
\def\BBpairsErr{3}  
\def\OffResLumi{22}  
\def\Yield{365} 
\def\YieldErr{49} 
\def\BRMean{10.2} 
\def\BRStat{1.4} 
\def\BRSyst{0.9} 
\def\AcpMean{-0.01} 
\def\AcpStat{0.13} 
\def\AcpSyst{0.02} 
\def\SigStat{8.7} 

\newcommand{\gevccc}{\ensuremath{{\mathrm{\,Ge\kern -0.1em V^2\!/}c^4}}\xspace}

\newcommand{\BaBarYear}      {06}
\newcommand{\BaBarNumber}    {059}
\newcommand{\SLACPubNumber} {12285}

\newcommand{\BaBarType}      {PUB}  

\begin{document}

\preprint{\babar-\BaBarType-\BaBarYear/\BaBarNumber}
\preprint{SLAC-PUB-\SLACPubNumber}

\title{Measurement of the {\boldmath{\Btorhopi\ }} Branching Fraction\\
  and Direct {\boldmath{\CP}} Asymmetry}

%
\author{B.~Aubert}
\author{M.~Bona}
\author{D.~Boutigny}
\author{F.~Couderc}
\author{Y.~Karyotakis}
\author{J.~P.~Lees}
\author{V.~Poireau}
\author{V.~Tisserand}
\author{A.~Zghiche}
\affiliation{Laboratoire de Physique des Particules, IN2P3/CNRS et Universit\'e de Savoie, F-74941 Annecy-Le-Vieux, France }
\author{E.~Grauges}
\affiliation{Universitat de Barcelona, Facultat de Fisica, Departament ECM, E-08028 Barcelona, Spain }
% ...
\collaboration{The \babar\ Collaboration}
\noaffiliation

\date{\today}

\begin{abstract}
We present improved measurements of the branching fraction and \CP\ asymmetry 
for the process \Btorhopi.  The
data sample corresponding to \LUMI~\ifb\ comprises \BBpairs\ million \FourS\to\BB\
decays collected with the \babar\ detector at the \pep2\ \abf\ at
SLAC.  The yield and \CP\ asymmetry are measured using an extended
maximum likelihood fitting method. The branching fraction and \CP\ 
asymmetry are found to be $ \BR(\Btorhopi) = [\BRMean~ \pm
\BRStat~\stat\ \pm \BRSyst~\syst ] \times 10^{-6}$
and $
\Acp(\Btorhopi) = \AcpMean~ \pm \AcpStat~\stat\ \pm{\AcpSyst}~\syst$.
\end{abstract}


\maketitle

Branching fraction and \CP asymmetry measurements of charmless \B\ 
meson decays provide valuable constraints for the determination of the
unitarity triangle constructed from elements of the
Cabibbo-Kobayashi-Maskawa quark-mixing
matrix~\cite{Cabibbo,KobayashiMaskawa}.  In particular, the angle
$\alpha \equiv
\arg\left[-V_{td}^{}V_{tb}^{*}/V_{ud}^{}V_{ub}^{*}\right]$ of the
Unitarity Triangle can be extracted from decays of the \B\ meson to
$\rho^\pm\pi^\mp$ final states~\cite{babarrhopi}.  However, the
extraction is complicated by the interference of decay amplitudes with
differing weak and strong phases.  One strategy to overcome this
problem is to perform an SU(2) analysis that uses all $\rho\pi$ final
states~\cite{QuinnAndSnyder}. Assuming isospin symmetry, the angle
$\alpha$ can be determined free of hadronic uncertainties from a
pentagon relation formed in the complex plane by the five
$\B\to\rho\pi$ decay amplitudes $B^0\rightarrow\rho^+\pi^-$,
$B^0\rightarrow\rho^-\pi^+$, $B^0\rightarrow\rho^0\pi^0$, $B^+\rightarrow\rho^+\pi^0$ and
$B^+\rightarrow\rho^0\pi^+$. These amplitudes can be determined
from measurements of the corresponding decay rates and \CP
asymmetries.  While all these modes have been measured~\cite{babarrhopi0,Bellerhopi0}, the current
experimental uncertainties need to be reduced substantially for a
determination of $\alpha$.  Here we present an update to previous
measurements of the 
\Btorhopi\ branching fraction and \CP asymmetry
$$\Acp=\frac{N(\Bm\to\rhom\piz)-N(\Bp\to\rhop\piz)}{N(\Bm\to\rhom\piz)+N(\Bp\to\rhop\piz)}.$$

The main additions compared to our previous 
analysis~\cite{babarrhopi0} are a larger dataset, a study of possible
backgrounds from higher $\rho$ resonances and the use of the $\rho$
mass in the maximum likelihood fit.

The data were collected with the \babar\
detector~\cite{BabarDet} at the \pep2\ asymmetric-energy \epem\
storage ring at SLAC.  Charged-particle trajectories are measured by a
five-layer double-sided silicon vertex tracker and a 40-layer drift
chamber located within a 1.5-T magnetic field.  Charged hadrons are
identified by combining energy-loss information from tracking (\dedx)
with the
measurements from a ring-imaging Cherenkov detector.  Photons are
detected by a CsI(Tl) crystal electromagnetic calorimeter with an
energy resolution of $\sigma_E/E=0.023(E/\gev)^{-1/4}\oplus 0.014$.
The magnetic flux return is instrumented for muon and \KL\
identification.  The data sample includes
$\BBpairs\pm\BBpairsErr$~million \BB\ pairs collected at the \FourS
resonance, corresponding to an integrated luminosity of \LUMI~\ifb.
 In addition, \OffResLumi~\ifb\ of
data collected 40~\MeV\ below the \FourS resonance mass are used for
background studies.  We perform full detector Monte Carlo (MC)
simulations equivalent to 460 \ifb\ of generic \BB\ decays and 140
\ifb\ of continuum quark-antiquark events 
($e^+e^- \rightarrow \qqbar,$\ $q=u,d,s,c$).  In addition, we
simulate over 50 exclusive charmless \B\ meson decay modes, including 1.4
million signal \Btorhopi\ decays.

\B\ meson candidates are reconstructed from
one charged track and two neutral pions.  The charged track used to
form the \Btorhopi\ candidate is required to have at least 12 hits in
the drift chamber, to have a transverse momentum greater than
0.1\gevc, and to be consistent with originating from the beam-spot.
It must have ionization-energy loss 
and Cherenkov angle signatures consistent with those expected for a pion. 
We remove charged tracks that pass electron
selection criteria based on \dedx\ and calorimeter information.
Neutral pion candidates are formed from two photon candidates, each with a
minimum energy of 0.03\gev\ and which are required 
to exhibit a lateral profile of energy deposition in the electromagnetic calorimeter
consistent with an  electromagnetic shower~\cite{BabarDet}.
The angular acceptance of photon candidates is restricted to exclude parts of
the calorimeter where showers are not fully contained.  We require the
photon clusters forming the \piz\ to be separated in space, with a
\piz\ energy of at least 0.2\gev\ and an invariant mass between 0.10
and 0.16\gevcc. 

Two kinematic variables, $ \Delta E = E^{*}_{B} -
\sqrt{s}/2$ and the beam-energy substituted mass of the \B\ meson $
\mes = \sqrt{(s/2 + {\bf p}_{0}\cdot {\bf p}_{B})^{2}/E^{2}_{0} - {\bf
p}^{2}_{B}}$, are used for the final selection of events.  Here
$E^{*}_{B}$ is the energy of the \B\ meson candidate in the
center-of-mass frame, $E_{0}$ and $\sqrt{s}$ are the total energies of
the $\epem$ system in the laboratory and center-of-mass frames,
respectively;and  ${\bf p}_{0}$ and ${\bf p}_{B}$ are the three-momenta of
the $\epem$ system and the \B\ meson candidate in the laboratory frame,
respectively.  For correctly reconstructed
\rhopi\ candidates \DeltaE\ peaks at zero, while for final states with a
charged kaon, such as $\Btokstarpi$, $\DeltaE$ is shifted by approximately
80~\MeV\ on average.  Events are selected with $5.20<\mes<5.29$\gevcc\
and $|\DeltaE|<0.20$\gev.  The \DeltaE\ limits remove background
from two- and four-body \B\ meson decays with a small loss in signal
efficiency.

Continuum events are the dominant
background. To suppress this background, we select only those events
where the angle \thetabsph\ in the center-of-mass frame between the
sphericity axis~\cite{sphericity} of the \B\ meson candidate's decays
products and the sphericity
axis of the rest of the event satisfies
$|\cos\thetabsph| < 0.9$.  In addition, we construct a non-linear
discriminant, implemented as an artificial neural network (\ANN) that
uses three input parameters: the zeroth- and second-order Legendre
event shape polynomials $L_0$ and $L_2$ calculated from the momenta
and polar angles, with respect to the \B\ meson thrust axis,  
of all charged particle and photon candidates not
associated with the \B\ meson candidate, and the output of a
multivariate, non-linear \B\ meson candidate flavor tagging
algorithm~\cite{BTagRef}. The output \ANN\ of the artificial neural
network peaks at 0.5 for continuum-like events and at 1.0 for \B\ 
meson decays. We require $\ANN>0.63$ which reduces the continuum background
by half for a 5\% loss in signal MC efficiency. To further improve the
signal-to-background ratio we restrict the invariant mass of the
$\rho$ candidate to $0.55 < m_{\pi\pi} < 0.95$\gevcc.

The average \B\ meson candidate multiplicity per event is 1.8 as 
neutral and charged pion combinatorics can lead to more than one \B\ meson
candidate. We choose the best candidate based on a \chisq\
formed from the measured masses of the two \pizero\ candidates within
the event compared to the known \pizero\ mass~\cite{pdg}. In the case
of multiple charged pion candidates the choice is random so as not to
bias the fit distributions. This random selection has a negligible impact on
 the systematic uncertainty. The total
\Btorhopi\ selection efficiency is $15.4\pm0.1\%$.  
In signal MC studies, the candidate is correctly reconstructed 54.9\%
of the time. The remaining candidates come from self-cross-feed (SCF,
37.5\%) and mistag events (7.6\%). SCF events stem primarily from
swapping the low energy \piz from the resonance with another from the
rest of the event. Signal events reconstructed with the wrong charge
are classified as mistag events.  Both SCF and mistag events emulate
signal events, however the resolution in \mes\ and
\DeltaE\ tends to be worse.

We use MC events to study the backgrounds from other \B-meson decays.
The dominant contribution comes from $b \rightarrow c$ transitions;
the next most important is from charmless \B-meson decays.  Seventeen
individual charmless modes show a significant contribution once the
event selection has been applied.  These modes are added into the fit
(described below) fixed at the yield and asymmetry determined by the
simulation, based on their measured values~\cite{pdg}.  The largest contributions
come from $\Bz \ra \rhopm \rhomp$ and $\Bz \ra \rhopm \pimp$.  For
$\Bz \ra \eta^\prime\piz$ and $\Bz \ra \Kstarpm \rhomp$ we use half
the measured upper limit~\cite{pdg}.  We estimate the $\Bz \ra
a_1^0\piz$ branching fraction from that of $\Bz \ra
a_1^+\pim$~\cite{a1pi0} using isospin relations.  If no charge
asymmetry measurement is available, we assume zero asymmetry.

Although all other states that decay like the $\rho$ to $\pi\piz$ --
the $\rho(1450)$ and the $\rho(1700)$, subsequently referred to
collectively as \rhostar\ -- lie outside our $\rho(770)$ mass cut, a
contribution to our signal cannot be  ruled out \emph{a priori}.  To
account for the possible presence of these modes, 
an unbinned maximum likelihood fit to the
\Btorhostarpi\ yield is performed in a sideband of the $m_{\pi\pi}$ invariant
mass. This fit uses the same algorithm as described below but with
only the three input variables \mes, \DeltaE, and \ANN. The mass
window is chosen to be as far as possible from the $\rho(770)$ mass,
centered near the pole of the $\rho(1700)$ at $1.5 < m_{\pi\pi} <
2.0$~\gevcc.  The fitted yield for the $B^\pm\ra\rho^{*\pm}\piz$ decay
is then extrapolated into the $\rho(770)$ region, $0.55 < m_{\pi\pi} <
0.95$\gevcc, using a non-relativistic
Breit-Wigner line-shape. Although the choice of mass range is motivated by the
$\rho(1700)$, any yield seen is attributed entirely to the
$\rho(1450)$, which is the closer of the two resonances to the
signal. From the $B^\pm\ra\rho^\pm(1450)\piz$ MC, the ratio of the
number of candidates in the sideband to candidates in the signal mass
region is approximately 12.6:1. The fit in the sideband yields $101
\pm 32$ events, resulting in an estimate of the $\rho^*$ background of
$8$ events. We investigate possible interference effects by using an analytical
model for the lineshapes of the $\rho(770)$ and the $\rho^*$. We
compare the use of relativistic and non-relativistic Breit Wigner
lineshapes and vary the widths of the lineshapes by their
uncertainties~\cite{pdg}. We also scan the relative phase between
the two resonances from $-\pi$ to $\pi$. 
We assign a conservative systematic uncertainty of
100\% for the $\rho^*$ background based on the largest change in the
number of events in the range $0.55 < m_{\pi\pi} < 0.95$\gevcc from
these tests. The $\rho^*$ then
enters into the nominal fit with PDFs constructed from
$\B^\pm\ra\rho^\pm(1450)\piz$ MC simulation.

The non-resonant \Btopipipi\ branching fraction has, to date, not been
measured. To estimate the size of its contribution we select a region
of the Dalitz plot --- defined by the triangle
$(m^2_{\pipm\piz_1},m^2_{\pipm\piz_2})=(6,6),(6,15),(11,11)$\gevccc\
--- that is far from the signal as well as the $\rho(1450)$ and higher
resonances and which has low levels of continuum background. The
unbinned maximum likelihood fit with only three input variables (\mes,
\DeltaE, and \ANN) is applied in this region. The only significant
backgrounds expected are from generic \B\ and continuum events.  The
yields of the generic \B\ decays are fixed to values expected from MC
simulation while the continuum and non-resonant yields are allowed to
float. There are 1100 data events in the selected Dalitz region and
the fit yields $-5.1\pm7.6$ non-resonant events. This is consistent
with zero and the non-resonant contribution is therefore not
considered as a background to our signal.

An unbinned maximum likelihood fit to the variables
\mes, \DeltaE, \ANN\ and $m_{\pi\pi}$ is used to extract
the total number of signal \Btorhopi\ and continuum background
events and their respective charge asymmetries.
The likelihood for the selected sample is given by the product
of the probability density functions (PDF) for each individual candidate,
multiplied by the Poisson factor:
$$ \Like = \frac{1}{N!}\,e^{-N^\prime}\,(N^\prime)^N\,\prod_{i=1}^N
{\cal P}_i\, ,$$ where $N$ and $N^\prime$ are the number of observed
and expected events, respectively. The PDF ${\cal P}_i$ for a given
event $i$ is a sum of the signal and background terms:
\begin{eqnarray}
  {\cal P}_i
  & = &
  N^{\rm Sig} \times  \frac{1}{2} \,
  [\,(1-Q_i A^{\rm Sig}) f_{\rm Sig}\, {\cal P}^{\rm Sig}_{i} 
   \nonumber
  \\
  &  &
    ~+~ (1 - Q_i A^{\rm Sig})\, f_{\rm SCF}\, {\cal P}^{\rm Sig}_{{\rm SCF},i}     
  \nonumber
  \\
  &  &
    ~+~ (1+Q_i A^{\rm Sig}) f_{\rm Mis}\, {\cal P}^{\rm Sig}_{{\rm Mis},i} 
  \,]
  \nonumber
  \\
  & & 
  + \sum _j N^{\rm Bkg}_{j}\times  \frac{1}{2}  (1 - Q_iA^{\rm Bkg}_{j})\,  {\cal
    P}^{\rm Bkg}_{j,i},
  \nonumber 
  \label{pdfsum}
\end{eqnarray}

\noindent where $Q_i$ is the charge of the pion in the event, $N^{\rm
Sig}(N^{\rm Bkg}_j)$ and $A^{\rm Sig}(A^{\rm Bkg}_j)$ are the yield
and asymmetry for signal (background) component $j$, respectively.
The fractions of true signal ($f_{\rm Sig}$), SCF signal ($f_{\rm
SCF}$), and wrong-charge mistag events ($f_{\rm Mis}$) are fixed to
the numbers obtained from MC simulations.  The $j$ individual
background terms comprise continuum, $b\to c$ decays, \rhostar, and
seventeen other exclusive charmless \B\ meson decay modes. 
Signal and continuum yields are allowed to float in the fit, with the 
generic \B\ yields fixed to values expected from MC simulation.  
The PDF for each component,
in turn, is the product of the PDFs for each of the fit input
variables, $ {\cal P} = {\cal P}(\mes,\DeltaE){\cal P}(\ANN){\cal
P}(m_{\pi\pi}).$ Due to correlations between \DeltaE\ and \mes, the
${\cal P}(\mes,\DeltaE)$ for signal and all background from \B\ meson decays
are described by two-dimensional non-parametric PDFs~\cite{2dkeys}
obtained from MC events.  For continuum background, ${\cal
P}(\mes,\DeltaE)$ is the product of two one-dimensional
non-parametric PDFs; \mes\ is well described by
an empirical phase-space threshold function~\cite{argus} and \DeltaE\
is parameterized with a second degree polynomial.  The parameters of
the continuum PDFs are allowed to float in the fit except for the endpoint of
the empirical phase-space threshold function which is fixed at
5.29\gevcc.  $\ANN$ is described by the product of an exponential and
a polynomial function for continuum background and by a Gaussian with
a power-law tail on one side~\cite{CrystalBall} for all other modes.
For ${\cal P}(m_{\pi\pi})$, one-dimensional non-parametric PDFs
obtained from MC events are used to describe all modes except the
signal mode itself, which is described by a non-relativistic
Breit-Wigner line-shape.  The parameters for this PDF are held fixed
to the MC values and varied within errors to estimate systematic
uncertainties.

A number of cross checks confirm that the fit is unbiased. 
Using a double Gaussian PDF instead of a Breit-Wigner or 
omitting $m_{\pi\pi}$ altogether as a fit variable has 
no significant effect on the measured branching fraction. In 1000 MC
pseudo-experiments, we use the maximum likelihood fit to
extract the yields and asymmetries.  The distributions for each
component are generated from the component's PDF, giving values for
the fit variables \mes, \DeltaE, \ANN\ and $m_{\pi\pi}$.  The expected
number of events is calculated from the branching fraction and
efficiency for each individual mode.  The generated number of events
for each fit component is determined by varying the expected
number according to a Poisson distribution.  The test is repeated
using samples with different asymmetry values.  We repeat these MC
studies using fully simulated signal \Btorhopi\ events instead of
generating the signal component from the PDFs.  This verifies that the
signal component is correctly modeled, including correlations between
the fit variables.  As another cross check we compare the distribution
of the helicity angle $\theta_{\rm Hel}$ between the momenta of the
charged pion and the \B\ meson in the $\rho$ rest frame in data with
that modeled in MC samples for a variety of selection criteria.
To investigate the possible effects of interference, we repeat the
analysis excluding events where both $m_{\pi^{\pm}\pi^0}$ combinations
were in the range 0.55 to 0.95 \gevcc; the branching fraction
decreases by 0.1\%. 
k

\renewcommand\baselinestretch{1.3}
\begin{table}[hbt]
  \begin{center}
  \caption[Systematic uncertainties]{Summary of the systematic uncertainties.}
  \begin{ruledtabular}
    \begin{tabular}{ll}
      \multicolumn{2}{c}{Absolute uncertainties on yields}\\ 
      Source                   &  $\sigma_{\rm Syst.}^{\rm Yield}$ (Events)\\
      \hline
      \B\ background normalization & $^{+\;\, 6.9}_{-\;\, 7.2}$ \\ 
      PDF shapes               & $^{+\;\, 4.7}_{-\;\, 4.2}$ \\ 
      SCF fraction             &  $\pm  12.2$ \\ 
      Mistag fraction          &  $\pm\;\,  2.0$ \\ 
      \DeltaE\ shift          &  $\pm\;\, 2.6$  \\ 
      \hline
      Total                    & $\pm 15$ \\ 
      \hline\hline
      \multicolumn{2}{c}{Relative uncertainties on \BR(\Btorhopi) }\\
      Source            &  $\sigma_{\rm Syst.}^{\BR} (\%) $ \\
      \hline
      Efficiency estimation        &  $\pm 7.3 $ \\ 
      \B\ counting                 &  $\pm 1.1 $ \\
      \hline
      Total      & $\pm 7.4$ \\
      \hline\hline
      \multicolumn{2}{c}{Uncertainties on \Acp }\\
      Source            &  $\sigma_{\rm Syst.}^{\Acp }$ \\
      \hline
      Background normalization      &  $\pm 0.006$ \\ 
      Background asymmetry          &  $\pm 0.024$ \\ 
      PDF shapes                    &  $\pm 0.001$ \\
      \hline
      Total                         &  $\pm 0.02$  \\
    \end{tabular}
  \end{ruledtabular}
  \label{tab:systematics}
  \end{center}
\end{table}
\renewcommand\baselinestretch{1.0}

Individual contributions to the systematic uncertainty are summarized
in Table \ref{tab:systematics}.  
For
each contributing exclusive \B\ meson decay mode, we vary the number of
events in the fit by its measured uncertainty, or by $\pm 100\%$ if
derived from an upper limit.  For the $b\to c$ component, we fix the
rate based on the number calculated from MC samples and vary the
amount based on the statistical uncertainty on this number.  The
shifts in the fitted yields are calculated for each mode in turn and
then added in quadrature to find the total systematic effect.  To take
into account the variation of the two-dimensional non-parametric PDFs
used for \DeltaE\ and \mes, we smear the MC-generated distributions
from which the PDFs are derived.  This is effectively done by varying
the kernel bandwidth~\cite{2dkeys} up to twice its original value.
For $m_{\pi\pi}$ and $\ANN$, the parameterizations determined from
fits to MC events are varied by one standard deviation. 
The systematic uncertainties are determined using the
altered PDFs and fitting to the final data sample. The overall shifts
in the central value are taken as the size of the systematic
uncertainty. We vary the SCF fraction by a conservative estimate of
its relative uncertainty ($\pm10\%$) and assign the shift in the
fitted number of signal events as the systematic uncertainty of the
SCF fraction.  
To account for differences in the neutral particle
reconstruction between data and MC simulation, the signal PDF
distribution in \DeltaE\ is offset by $\pm 5 \mev$ and the data are
then refit.  The larger of the two shifts in the central value of
the yield is 2.6 events, which is taken as the systematic uncertainty
for this effect.


\begin{figure}[htb]
  \renewcommand\baselinestretch{0.5}
  \begin{tabular}{cc}
    \epsfig{file=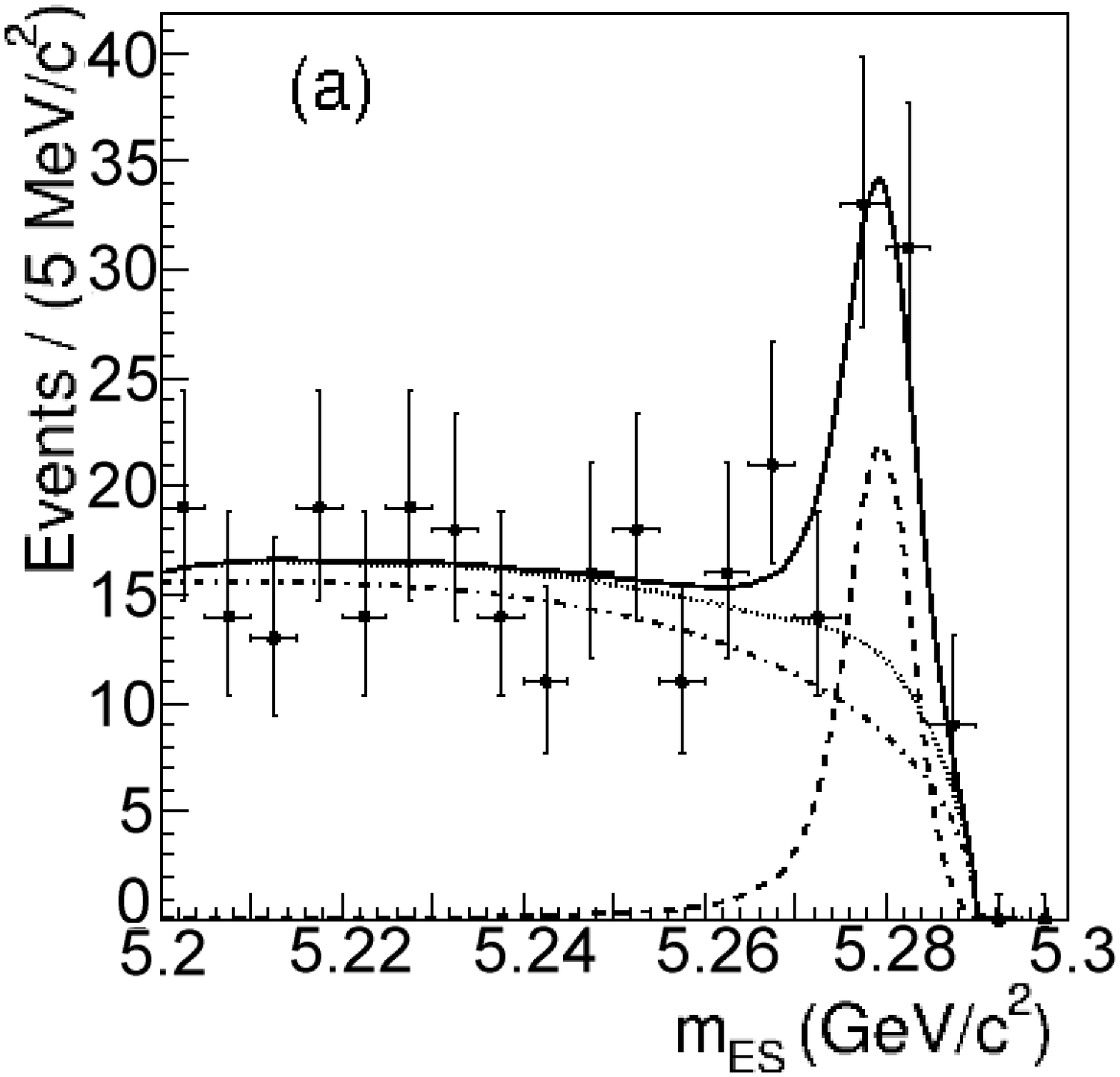,width=0.51\columnwidth}
    &
    \epsfig{file=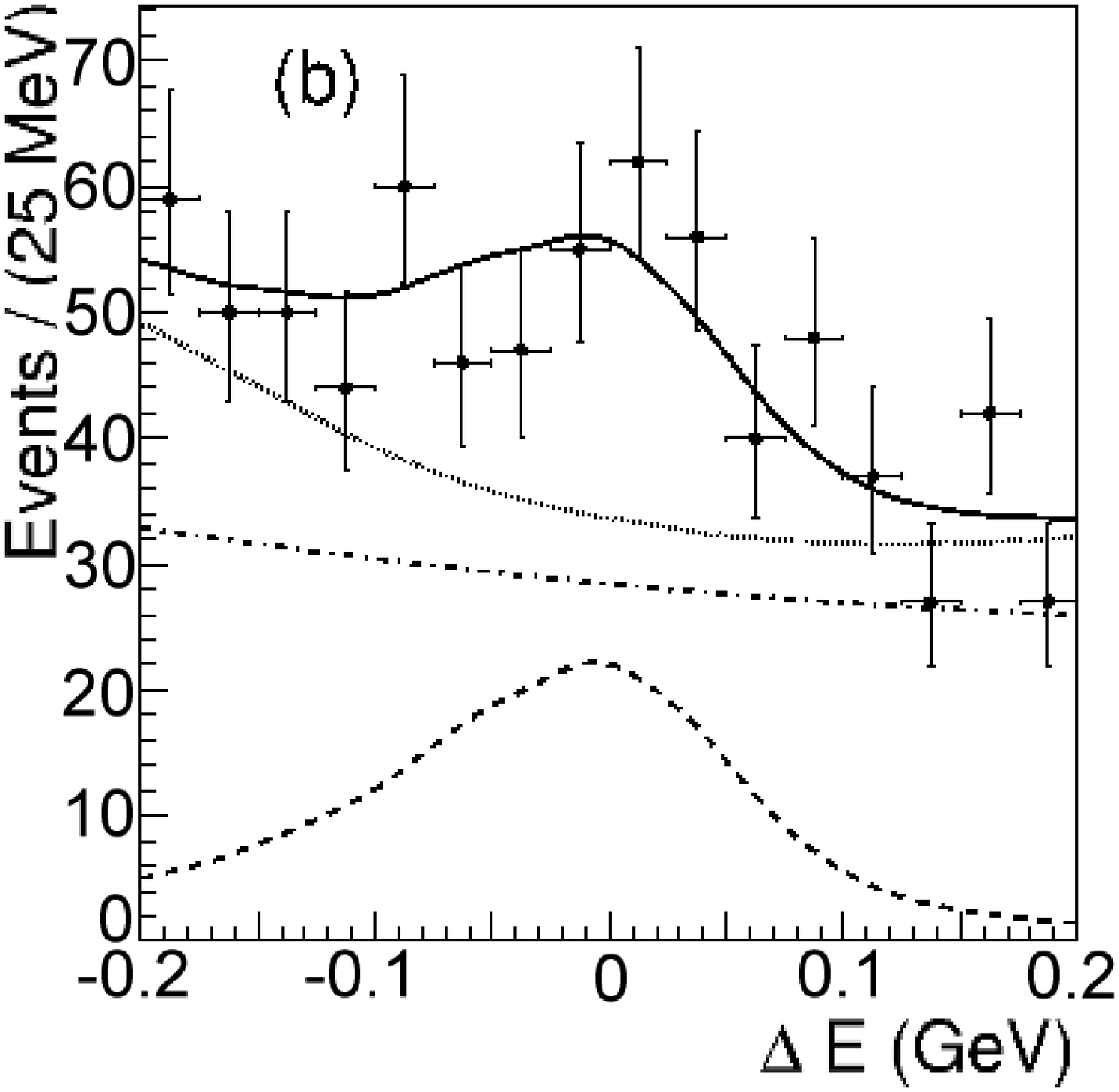,width=0.51\columnwidth}
    \\
    \epsfig{file=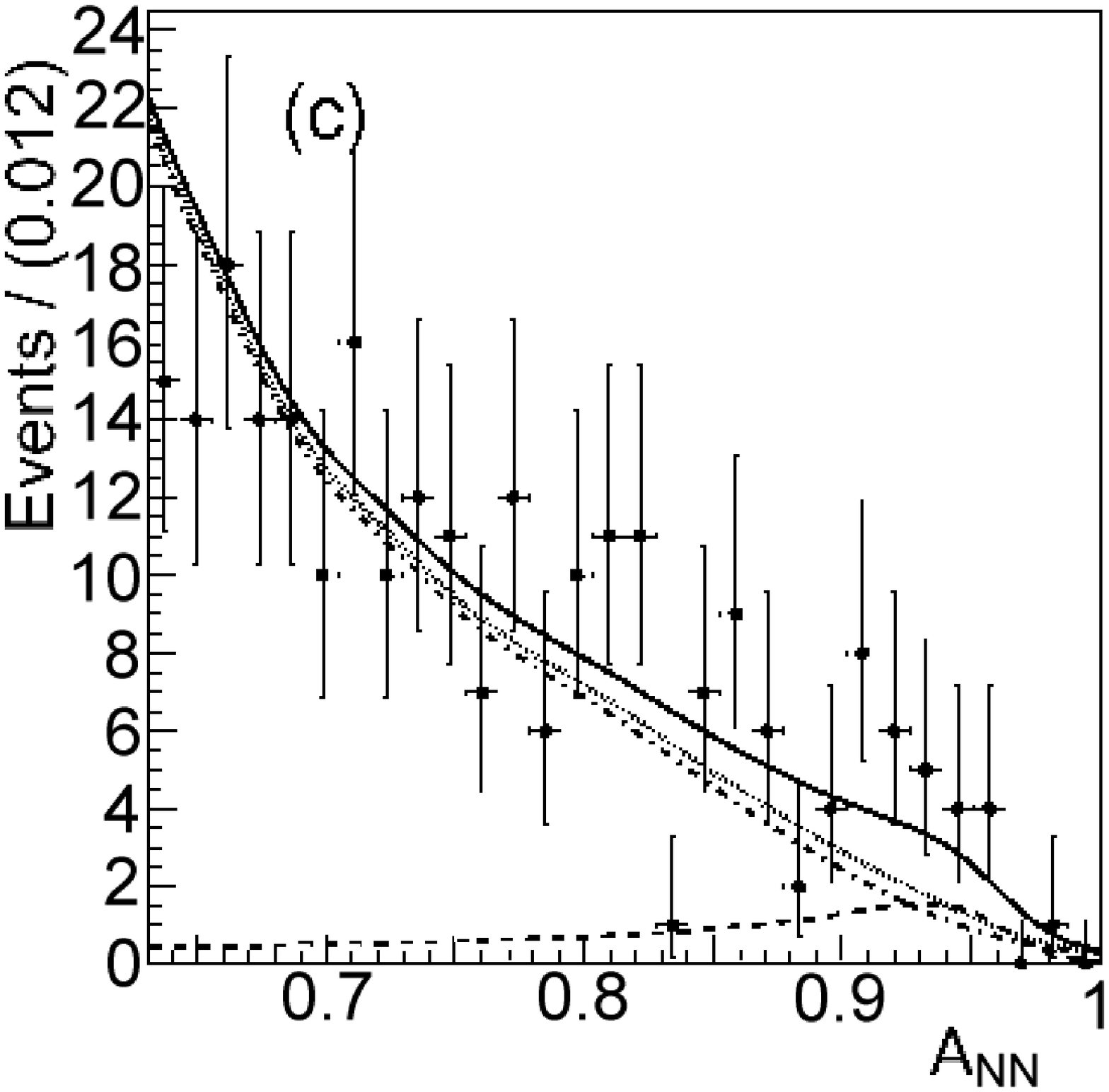,width=0.51\columnwidth}
    &
    \epsfig{file=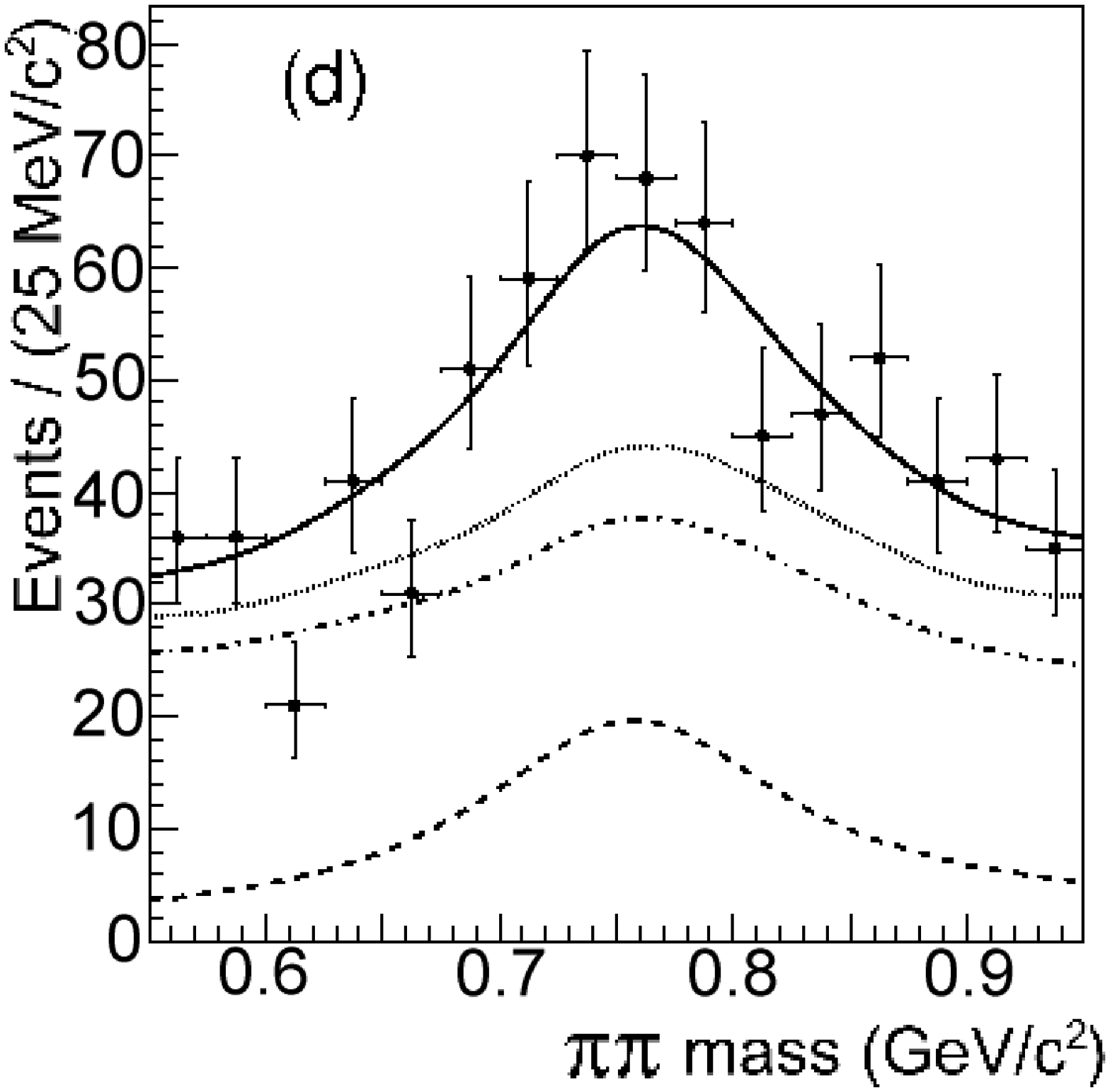,width=0.51\columnwidth}
    \\
  \end{tabular}
  \renewcommand\baselinestretch{1.0}
  \caption[Results for the four variables of the maximum likelihood fit]
  {Likelihood projection plots for the four fit variables,
    (a) \mes, (b) \DeltaE, (c) \ANN, and (d) $m_{\pi\pi}$.
    Each plot shows the total PDF (solid line),
    total background (dotted line),
    continuum contribution (dotted-dashed line),
    and the signal component (dashed line).
}
  \label{fig:results}
\end{figure}

Corrections to the \pizero\ energy resolution and efficiency,
determined using various data control samples, add a systematic uncertainty
of 7.2\%.  A relative systematic uncertainty of $1\%$ is assumed for
the pion identification.  A relative systematic uncertainty of 0.8\%
on the efficiency for a single charged track is applied.  Adding all
the above contributions in quadrature gives a relative systematic
uncertainty on the branching fraction of 7.3\%.  Another contribution
of $1.1\%$ comes from the uncertainty on the total number of \B\ events.  

To calculate the effects of systematic shifts in the charge
asymmetries of background modes, the asymmetry of 
each mode is varied by its measured
uncertainty.  For contributions with no asymmetry measurement, we assume zero
asymmetry and assign an uncertainty of 20\%, motivated by the largest
charge asymmetry measured in any mode so far~\cite{Bkpi2004}.  The
individual shifts are then added in quadrature to find the total
systematic uncertainty.  In addition, the effect of altering the
normalizations of the \B\ backgrounds affects the fitted asymmetry.
The size of the shift on the fitted \Acp\ is taken as the size of the
systematic uncertainty.

The central value of the signal yield from the maximum likelihood fit
is $\Yield\pm\YieldErr$ events, with a background of $44840\pm217$ continuum events
and an expected background of $842\pm 34$ events from other \B\ decays.
Fig.~\ref{fig:results} shows the distributions of $\mes$, $\DeltaE$,
\ANN\ and $m_{\pi\pi}$.  
The plots are enhanced in signal by selecting only those events which
exceed a threshold of 0.1 (0.05 for \ANN) for the likelihood
ratio~\cite{Kstarpi} $R = (N^{\rm Sig}{\cal P}^{\rm Sig})/(N^{\rm
Sig}{\cal P}^{\rm Sig} +
\sum_{i} N^{\rm Bkg}_{i}{\cal P}^{\rm Bkg}_{i})$,
where $N$ are the central values of the yields from the fit and ${\cal
P}$ are the PDFs with the projected variable integrated out.  This
threshold is optimized by maximizing the ratio $S = (N^{\rm
Sig}~\epsilon^{\rm Sig})/\sqrt{N^{\rm Sig}~\epsilon^{\rm Sig} +
\sum_{i} N^{\rm Bkg}_{i}~\epsilon^{\rm Bkg}_{i}}$ where $\epsilon$
are the efficiencies after the threshold is applied. The PDF
components are then scaled by the appropriate $\epsilon$. The
efficiencies for the likelihood ratios vary for each
variable and result in a different number of events in each
projection. Compared against the null
hypothesis, the statistical significance $\sqrt{-2 \ln
(\Like_{Null}/\Like_{Max})}$ of the signal yield amounts to
\SigStat\ standard deviations. We obtain
$\BR(\Btorhopi) = [\BRMean \pm \BRStat \pm \BRSyst]\times 10^{-6}$,
and $ \Acp = \AcpMean \pm {\AcpStat} \pm {\AcpSyst}$,
where the first error is statistical and the second error systematic.
The measurements are consistent with previous
results~\cite{babarrhopi0} and provide improved constraints for the
determination of the angle $\alpha$ from $B\to \rho\pi$ decays.

We are grateful for the excellent luminosity and machine conditions
provided by our \pep2\ colleagues, 
and for the substantial dedicated effort from
the computing organizations that support \babar.
The collaborating institutions wish to thank 
SLAC for its support and kind hospitality. 
This work is supported by
DOE
and NSF (USA),
NSERC (Canada),
IHEP (China),
CEA and
CNRS-IN2P3
(France),
BMBF and DFG
(Germany),
INFN (Italy),
FOM (The Netherlands),
NFR (Norway),
MIST (Russia), and
PPARC (United Kingdom). 
Individuals have received support from CONACyT (Mexico), A.~P.~Sloan Foundation, 
Research Corporation,
and Alexander von Humboldt Foundation.

\end{document}